\documentclass{IEEEtran}
\usepackage{cite}
\usepackage{amsmath,amssymb,amsfonts}
\usepackage{mathtools}
\usepackage{cuted}
\usepackage{enumitem}
\usepackage{algorithmic}
\usepackage{graphicx}
\usepackage{textcomp}
\usepackage{xcolor}
\def\BibTeX{{\rm B\kern-.05em{\sc i\kern-.025em b}\kern-.08em
    T\kern-.1667em\lower.7ex\hbox{E}\kern-.125emX}}
\begin{document}
\title{A Reconfigurable Intelligent Surface With Surface-Wave Assisted Beamforming Capabilities}
\author{Vasileios~G.~Ataloglou, \IEEEmembership{Member, IEEE}, and George~V.~Eleftheriades, \IEEEmembership{Fellow, IEEE}
\thanks{Submitted on 03/10/2024.}
\thanks{V.~G.~Ataloglou and G.~V.~Eleftheriades are with the Edward S. Rogers Sr. Department of Electrical \& Computer Engineering, University of Toronto, Toronto ON M5S3G4, Canada (e-mail: vasilis.ataloglou@mail.utoronto.ca, gelefth@waves.utoronto.ca).}
}

\maketitle

\begin{abstract}
The integration of tunability mechanisms in the metasurface design has unleashed a tremendous potential for wireless communications. In particular, reconfigurable intelligent surfaces (RISs) can manipulate the reflections of an incident electromagnetic wave at will, based on the real-time conditions, with the aim of enhancing communication links. In this paper, we develop an RIS at C band that operates with a transverse-electric (TE) polarization and can shape the radiation pattern at a single plane with high-accuracy, in addition to the more conventional beamsteering functionalities. The beamforming is facilitated  by subwavelength unit cells that allow the excitation of auxiliary surface waves in the vicinity of the RIS. Importantly, these evanescent fields are predicted and harnessed through an integral-equation framework used for the analysis and optimization of the RIS. A fabricated prototype demonstrates beamsteering up to $\pm60^\circ$ with an average illumination efficiency of $95\%$ and sector patterns with a varying beamwidth (ranging from $30^\circ$ to $60^\circ$) that verify the full-wave simulations. Lastly, losses •are predicted and constrained during the optimization stage leading to solutions with relatively high power efficiency.
\end{abstract}


\section{Introduction}
\label{sec:introduction}
The continuously growing number of interconnected devices requiring constant, reliable and high data-rate connectivity poses challenges to the development of future generations (6G and beyond) of wireless communication networks. Among the innovative ideas to overcome these challenges, reconfigurable intelligent surfaces (RISs) have been proposed as a way to modify the propagation environment and improve communication links between users. An RIS is a tunable metasurface consisting of subwavelength scatterers that can be programmed to anomalously reflect an incident electromagnetic wave towards one or more desired directions \cite{Basar:Access2019,ElMossalllamy:TCCN2020, Khalily:ICM2022}. In this way, RISs are utilized to increase the power of the received signals in areas that are not sufficiently covered by the transmitter due to obstacles or multipath interference. Moreover, RISs are passive structures acting directly on the radio-frequency (RF) signal and they usually require only a DC external bias \cite{DiRenzo:OJCS2020}. Therefore, they offer a cost-effective and power-efficient solution for the growing requirements of wireless communications.

The hardware implementation of reconfigurable metasurfaces operating at microwaves has been thoroughly investigated and different tunable elements and design methods have been proposed \cite{Sievenpiper:TAP2003, Hum:TAP2007, Cui:TAP2014, Dai:Access2020, Yang:TAP2016, Araghi:Access2022, Gros:OJCS2021, Trichopoulos:OJCS2022, Popov:AoM2021, Bagheri:Access2023, Hu:AWPL2023}. In the vast majority of cases, the RIS is designed as a tunable reflectarray with each unit cell inducing locally a spatially-varying phase in the reflected field. In the case that PIN diodes are used, the reflected phase takes discrete values, and the structure is typically called an $n$-bit RIS ($n$ being the number of diodes per cell for a single polarization) \cite{Yang:TAP2016, Han:AWPL2019, Yin:TAP2024}. On the other hand, varactor diodes offer a continuous tunability of the reflecting phase, avoiding any drop in efficiency due to phase quantization. Additionally, varactor diodes operate under a reverse DC bias voltage and, therefore the DC power dissipation is minimal. These are the reasons that varactor diodes were used in our presented work. Lastly, for an RIS application, the metasurface is typically placed in the far-field region of the transmitting antenna, rendering the incident illumination approximately a uniform plane wave. This is in contrast to the earlier implementations of such tunable reflective surfaces that were focused on the realization of reconfigurable reflectarray antennas with a feed placed in the vicinity of the metasurface \cite{Hum:TAP2007}.

For RIS applications, it is often important to modify not only the direction of the main beam, but also the beam shape. For instance, if a wider angular range needs to be covered by a single RIS, a sector pattern of larger beamwidth may be preferable compared to a narrower highly-directive beam. While several RIS studies focus on efficient beamsteering of the reflected beam, the beamshaping capabilities of the RIS and the design methods to realize them are often overlooked. A beamshaped far-field pattern would require both amplitude and phase tapering throughout the metasurface aperture. One way to fulfill this requirement is to deliberately introduce ohmic losses that are controlled independently from the reflected phase, thus providing a way to acquire the required complex reflection coefficients \cite{Demetre:TAP2023,Kossifos:TAP2024}. However, this approach limits the power efficiency, especially in cases that a low reflection amplitude is required for a large portion of the unit cells. Alternatively, optimization techniques based on a reflect-array approach may be used to optimize the reflected phases including the coupled loss to achieve some constrained beamshaping effects \cite{Bagheri:Access2023}. Yet, the calculation of the far-field based on the local reflection coefficient at each cell usually disregards mutual coupling and leads to deviations from the expected pattern. In particular, these discrepancies are significant when the unit cell size is smaller than the typical choice of half-wavelength and surface waves are excited by means of fast-varying parameters between adjacent cells. Metallic vias acting as baffles have been examined as a way to eliminate coupling, but the approach can only be applied for a single polarization and increases the fabrication complexity~\cite{Xu:TAP2019,Liang:TAP2022}.

As another option, integral equation methods can be utilized to optimize an aperiodic finite metasurface for desired far-field patterns \cite{Budhu:TAP2020, Xu:Access2022, Ataloglou:APS2022_RIS}. In this case, the mutual coupling between different cells is accounted for and patterns can be realized through a rigorous optimization framework with higher precision compared to traditional array factor approaches. In fact, the interactions between cells can be harnessed to excite auxiliary surface waves that allow for amplitude tapering without deliberately introducing losses \cite{Epstein:PRL2016}. Specifically, surface waves can facilitate wave transformations in reflection and transmission with theoretically unity power efficiency, that is limited in practice only by the materials and lossy parts of the metasurface \cite{Epstein:PRL2016,Kwon:PRB2018,Ataloglou:AWPL2021}. Therefore, the benefits of this design method lie both on the more accurate modeling of the metasurface, and on the possibility to achieve higher power efficiency for patterns requiring amplitude tapering through the excitation of surface waves.

An RIS has been recently optimized based on an integral equation framework in \cite{Popov:AoM2021}. Compared to that work, we formulate the integral equations based on analytical expressions without the need of preliminary simulations to obtain the Green's function that depends on the substrate parameters and the locations of loaded wires. The use of analytical expressions facilitates the design process, as it is much faster to modify the density of loaded wires and substrate parameters, until a desirable performance is obtained. Moreover, we incorporate the unit cell losses into the framework and try to minimize them during optimization to achieve the highest-possible power efficiency, while maintaining the desired far-field radiation. Lastly, the beamsteering and beamshaping capabilities of the RIS are explored and verified experimentally. While the presented RIS is biased uniformly along one dimension, limiting its beamforming capabilities to a single plane and TE-polarized fields, the design framework can be expanded to RISs varying in both dimensions, by modifying the Green's functions \cite{Ataloglou:APS2024}, and individually biasing all unit cells.

The rest of the paper is organized as follows. Section~\ref{sec:MoM} describes the homogenized model used for the design and optimization of the RIS. The unit cell topology is described in Sec.~\ref{sec:cell}, while simulation results and a comparison with traditional approaches to design RISs are given in Sec.~\ref{sec:simulation}. The measurements of the fabricated prototype are discussed in Sec.~\ref{sec:measured} and conclusions are made in Sec.~\ref{sec:conclusion}. Lastly, some technical details for the extraction of the performance metrics are detailed in the Appendix. It is noted that our work focuses on the design framework and electromagnetic performance of the presented reconfigurable metasurface, while it disregards any communication-related aspects and requirements of RISs (e.g., channel estimation, modulation schemes) that are often discussed in theoretical communication-focused works \cite{Yang:TranCommun2020,Jiang:JSAC2021}. Yet, the term reconfigurable intelligent surface (RIS) is adopted herein due to its wide use in similar works focusing also on the reconfigurable manipulation of the reflections from an incident electromagnetic wave.

\section{Analysis and Optimization Framework}
In this section, the framework to model, analyze and optimize the RIS is presented. The integral-equation approach solved with a Method of Moments is based on a previous work presented in \cite{Xu:Access2022} for static metasurfaces illuminated by an embedded source, but the main steps are included herein for completeness.
\label{sec:MoM}
\subsection{Geometry and simplified model}
\begin{figure}
\centering
\includegraphics[width=0.8\columnwidth]{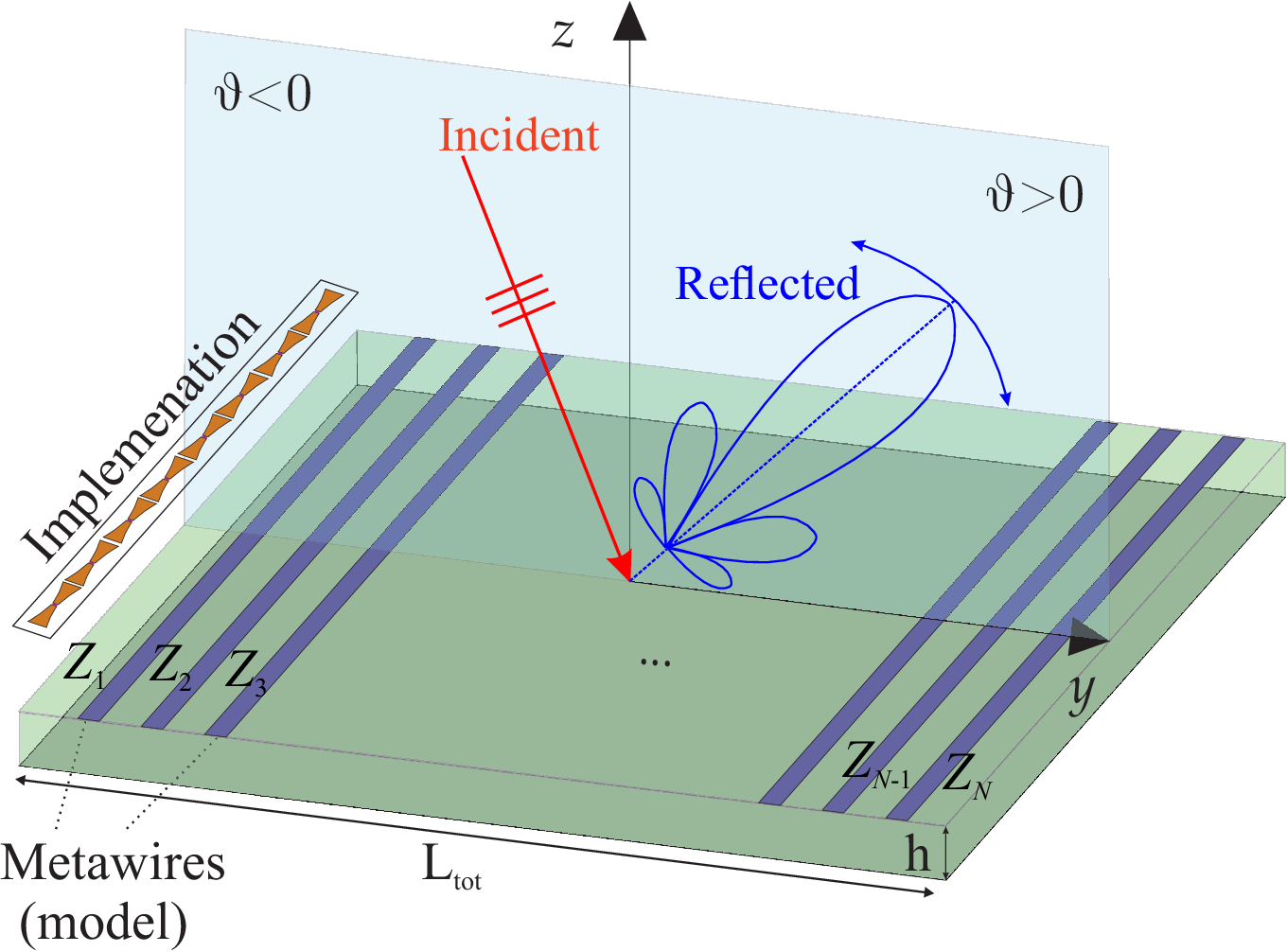}
\caption{\label{fig:Sketch} Sketch of the RIS consisting of $N$ rectangular sheets of homogenized impedances $Z_n$ along a length of $L_\mathrm{tot}$. Each homogenized sheet is implemented with varactor-loaded cells that are repeated periodically along the $x$-axis.}
\end{figure}

As sketched in Fig.~\ref{fig:Sketch}, the RIS consists of an impedance layer etched on a grounded substrate of thickness $h$, dielectric constant $\varepsilon_r$ and loss tangent $\mathrm{tan}(\delta)$. The impedance layer incorporates the unit cells, that are loaded with varactors and repeat periodically along the $x$-axis. In turn, the effect of the unit cells is homogenized into rectangular sheets of width $w$ and uniform surface impedance $Z_n=R_n+jX_n (n=1,...,N)$, where $N$ is the number of impedance strips. It is noted that the RIS has a length $L_\mathrm{tot}$ in the $y$-direction, while it is considered uniform along the $x$-axis, i.e. the biasing is identical for the cells corresponding to the same impedance strip. Furthermore, the RIS is illuminated by a TE-polarized wave ($\mathbf{E}^\mathrm{inc}=E^\mathrm{inc} \mathbf{\hat{y}}$) that induces $y$-polarized currents. The simplified model allows the fast and highly-accurate calculation of the reflected far-field pattern through integral-equations for any set of surface impedances $Z_n$ that corresponds to a set of bias voltages $V_n$. In turn, this enables the optimization of the applied DC bias voltages $V_n$ for desired far-field patterns.

\subsection{Volume-surface Integral Equations}  \label{subsec:VSIE}
When the RIS is illuminated by an incident wave, surface current densities will be induced on the impedance sheets and the ground plane, denoted by $J_w(y)$ and $J_g(y)$, respectively. In addition, the effect of the dielectric is considered through an induced volumetric current density $J_v(y,z)$ that takes nonzero values within the substrate region. The free-space two-dimensional (2D) Green's function is utilized to calculate the scattered fields as:
\begin{align}\label{eq:scat}
E_{i}^\mathrm{sc} (\text{\boldmath$\rho$})= \begin{dcases}
-\frac{k \eta}{4} \int_{C_i} J_i (y') H^{(2)}_0 (k|\text{\boldmath$\rho$}-\text{\boldmath$\rho'$}|) dl', \ i=\{w,g\}, \\
-\frac{k \eta}{4} \int_{S_i} J_i (y',z') H^{(2)}_0 (k|\text{\boldmath$\rho$}-\text{\boldmath$\rho'$}|) ds', \  \  \  \ i=v,
\end{dcases}
\end{align}
where $C_w,C_g,S_v$ refer to the domains of the induced currents on the impedance sheets, ground plane and dielectric, respectively. In addition, $k$ is the free-space wavenumber, $\eta \approx 377 \Omega$ is the free-space impedance, $ H^{(2)}_0$ is the second-kind Hankel function of zeroth order, and $\text{\boldmath$\rho$}=y\mathbf{\hat{y}}+z\mathbf{\hat{z}}, \text{\boldmath$\rho'$}=y'\mathbf{\hat{y}}+z'\mathbf{\hat{z}}$ are the position vectors of the observation and source points, respectively.

The total field, which is $x$-polarized, should be zero on the ground plane based on the boundary conditions, since the latter is considered a perfect electric conductor. Moreover, the impedance boundary condition, formulated by Leontovich and Shchulkin \cite{Pelosi:APM1996}, requires that the ratio of the total electric field to the induced current density is $Z_n$ at the $n$-the impedance strip \cite{Senior:ImpedanceBook1995}. Lastly, the polarization current is written as a function of the total electric field through the dielectric constant of the substrate. In summary, the total electric field satisfies:
\begin{align} \label{eq:VSIE}
E^\mathrm{tot} &=
E^\mathrm{inc}+E_w^\mathrm{sc}+E_g^\mathrm{sc}+E_v^\mathrm{sc} \nonumber \\ &= \begin{dcases}
Z(y) J_w(y), & \mathrm{on} \ C_w, \\
0, & \mathrm{on} \ C_g, \\
\frac{1}{j \omega (\varepsilon_r-1) \varepsilon_0} J_v(y,z), & \mathrm{in} \ C_v,
\end{dcases}
\end{align}
where $\omega$ is the angular frequency of operation and $Z(y)$ comprises the surface impedance values $Z_n$ at the location of each impedance strip. Equations~\eqref{eq:scat}-\eqref{eq:VSIE} define a system of volume-surface integral equations (VSIE) with respect to the unknown current densities.

The VSIE are solved with a Method of Moments (MoM) approach. Specifically, the surface current densities $J_w(y)$ and $J_g(y)$ are expanded to $N_w$ and $N_g$  1-D pulse basis functions extending along the impedance sheets and ground plane, respectively. Moreover, a total of $N_v$ 2-D pulse basis functions are utilized for the expansion of the volumetric current density $J_v(y,z)$. At each one of these basis functions the current is assumed constant with an unknown complex amplitude. By performing testing at the center of each basis function domain (point matching), a linear system of equations can be obtained with $(N_w+N_g+N_v)$ equations and unknowns, which can be written in a block-matrix format as follows:
\begin{align} \label{eq:MoM}
\begin{pmatrix}
& \mathbf{G}_{gg} & \mathbf{G}_{gv} & \mathbf{G}_{gw} \\
& \mathbf{G}_{vg} & \mathbf{G}_{vv}-\mathbf{P} & \mathbf{G}_{vw} \\
& \mathbf{G}_{wg} & \mathbf{G}_{wv} & \mathbf{G}_{ww}-\mathbf{Z}_{w}
\end{pmatrix}
\begin{pmatrix}
\bar{J}_g \\
\bar{J}_v \\
\bar{J}_w
\end{pmatrix} =
\begin{pmatrix}
\bar{E}_g^\mathrm{inc} \\
\bar{E}_v^\mathrm{inc} \\
\bar{E}_w^\mathrm{inc}
\end{pmatrix}.
\end{align}
In Eq.~\eqref{eq:MoM}, the vectors $\bar{J}_i$ and $\bar{E}_i^\mathrm{inc}$ ($i=\{w,g,v\}$ contain the unknown complex amplitudes of the current densities and the sampled incident field values, respectively, at the center of each basis function domain. The matrix $\mathbf{P}$ is diagonal with all elements being equal to $1/[j\omega (\varepsilon_r-1)\varepsilon_0]$, whereas the matrix $\mathbf{Z}_w$ is also diagonal and contains the surface impedance values corresponding to the location of each basis function along the impedance strips. Lastly, the matrices $\mathbf{G}_{ij}$ represent the self and mutual impedances with their elements $\mathbf{G}_{ij}[n][m]$ calculated analytically \cite{Xu:Access2022}.

By solving the linear system in Eq.~\eqref{eq:MoM}, the induced current densities are calculated for a given incident field and set of surface impedances $Z_n$. Then, the far-field radiation can be computed at a number of equally-spaced discrete angles $\theta_n$ in the $yz$ plane. By convention, we assume positive angles ($\theta_n>0$) for angles in the half-space with $y>0$, and negative angles ($\theta_n>0$) for the half-space with $y<0$. The field pattern is obtained from the sampled current densities as:
\begin{align}\label{eq:Eff}
\bar{E}^{ff}= \begin{bmatrix} \mathbf{G}_{fg} &  \mathbf{G}_{fv} & \mathbf{G}_{fw} \end{bmatrix} \begin{bmatrix}
\bar{J}_g \\
\bar{J}_v \\
\bar{J}_w
\end{bmatrix},
\end{align}
where $|\bar{E}^{ff}|$ is a vector containing the field pattern values in the discretized angles $\theta_n$ (the phase of $\bar{E}^{ff}$ is disregarded, as the focus is on the amplitude of the pattern), and the elements of the $\mathbf{G}_{fi}$ matrices are given as:
\begin{subequations}
\begin{itemize}[leftmargin=10pt]
\item $i=\{g,w\}$,
\begin{align} \label{eq:Gf_coefficients}
\hspace{-10pt} \mathbf{G}_{fi}[n][m]= -\frac{\eta \Delta_i}{4} \sqrt{\frac{2jk}{\pi}} \mathrm{exp} \{jk(y_m \mathrm{sin} (\theta_n) + z_m \mathrm{cos} (\theta_n))\},
\end{align}
\item $i=v$,
\begin{align}
\hspace{-10pt} \mathbf{G}_{fi}[n][m]=-\frac{\eta \pi r_0^2}{4} \sqrt{\frac{2jk}{\pi}} \mathrm{exp} \{jk(y_m \mathrm{sin} (\theta_n) + z_m \mathrm{cos} (\theta_n))\},
\end{align}
\end{itemize}
\end{subequations}
where the coordinates $(y_m,z_m)$ are the centers of the $m$-th current basis function. The radiation intensity is calculated at each angle $\theta_n$ as $U(\theta_n)=|E^{ff}(\theta_n)|^2/(2\eta)$. Then, the 2-D directivity (referring to the $yz$ cross-section) at each angle is given as:
\begin{align}\label{eq:Directivity}
D(\theta_n)=2\pi \frac{U(\theta_n)}{P_\mathrm{rad}},
\end{align}
where the radiated power $P_\mathrm{rad}$ is found through a numerical integration of the radiation intensity:
\begin{align}\label{eq:Prad}
P_\mathrm{rad}=\int_{-\pi/2}^{\pi/2} U(\theta) d\theta.
\end{align}
Equations~\eqref{eq:Eff}-\eqref{eq:Prad} are used to calculate the directivity in the far-field. However, measurements in the used bi-static measurement setup are taken in a finite radius $r_d$. Since a small difference in the pattern may exist, it may be beneficial to calculate the exact field without applying any far-field approximation. For each angle $\theta_n$, the total field $E^{nf}$ is calculated at the points $(y,z)=(r_{d}\mathrm{sin}(\theta_n),r_{d}\mathrm{cos}(\theta_n))$ by multiplying the weights $\bar{J}_i$ with the respective Green's functions. The radiation intensity is then calculated at this ``near-field" radius as $U(r_d,\theta_n)=r_d^2 |E^{nf}(\theta_n)|^2/(2\eta)$. Naturally, as $r_d \to \infty$, the calculation of the directivity at the radius $r_d$ converges to the far-field calculation.

\subsection{Optimization of surface impedances}
The analysis framework described above can be supplemented with an optimization technique to determine the surface impedances $Z_n$ that shape the reflections from the RIS into a desired far-field pattern. The optimization variables are the imaginary parts of the surface impedances $X_n$, with the real parts $R_n(X_n)$ added based on a lookup table for the specified unit cell. The synthesis problem is formed as the minimization of the cost function:
\begin{align} \label{eq:cost_function}
F=\sum_{n} \left|\frac{|E^{ff}(\theta_n)|}{\max\limits_n \{|E^{ff}(\theta_n)|\}} - \frac{|E^{ff}_\mathrm{des}(\theta_n)|}{\max\limits_n \{|E^{ff}_\mathrm{des}(\theta_n)|\}}\right|^2,
\end{align}
where $|E^{ff}_\mathrm{des}|$ is the desired field pattern. The expression in \eqref{eq:cost_function} quantifies the difference in the normalized field pattern values over all angles, and its minimization guarantees good matching between the obtained radiation pattern and the desired one.

Two types of constraints are included in the optimization problem. First, the imaginary part of the surface impedances are constrained within a range of realizable values $X_n \in [X_\mathrm{min},X_\mathrm{max}]$, as obtained from the unit cell simulation for the respective range of the varactor's capacitance. Secondly, supplementing the losses of the unit cell into the surface impedance through the real part $R_n$ allows to predict the total reflected power $P_\mathrm{rad}$. Thus, a constraint is included with regard to the minimum power efficiency $\eta_\mathrm{min}$:
\begin{align}\label{eq:e_p_min}
\eta_p=\frac{P_\mathrm{rad}}{P_\mathrm{min}} > \eta_\mathrm{min},
\end{align}
where $P_\mathrm{inc}$ is the total normal incident power to the RIS. As expected, if the value $\eta_\mathrm{min}$ is selected to be unrealistically high, the radiation pattern matching of the optimized solution degrades. Thus, the value $\eta_\mathrm{min}$ is modified gradually until a good compromise between power efficiency and pattern matching is achieved for each presented case. In general, the benefit of incorporating losses in the framework and constraining them through Eq~\eqref{eq:e_p_min} is to avoid convergence to any solutions that exhibit unnecessarily high losses.

The optimization consists of two steps. In the first one, the built-in genetic algorithm optimization in MATLAB is utilized to arrive at a quasi-random starting point. Subsequently, this starting point is provided to a gradient-descent optimization that converges to the final solution. Due to the randomness of the first step, the process is repeated $10$ times and the solution with the lowest value for the cost function is selected. For simple beamsteering cases, a starting point could have been defined analytically based on a phase gradient. On the other hand, for beamforming cases that an intuitive selection of a good starting point may not be possible, the genetic algorithm could have been used as the sole optimization method at the expense of higher computational time. The latter is the reason for utilizing the genetic algorithm with only few generations and the sole purpose of getting initial points for the faster gradient-descent optimization.

Solving for the induced currents at every optimization iteration is greatly accelerated by eliminating the unknowns $\bar{J}_g$ and $\bar{J}_v$ from Eq.~\eqref{eq:MoM} and formulating a linear system of size $N_w \times N_w$ that calculates solely the currents $\bar{J}_w$. This method was described in \cite{Xu:Access2022}, and it is usually referred to as Kron's reduction (or node elimination) in power systems analysis \cite{Grainger:PowerBook1994}. Additionally, providing a semi-analytical gradient for the cost function, as also suggested in \cite{Xu:Access2022}, speeds up the gradient-descent optimization, compared to using finite differences. A single repetition of the two-stage optimization for an RIS comprising $30$ meta-wires segmented into $5$ basis functions each ($N_w=150$) takes on average around $48$ seconds in a standard desktop computer.

\section{Unit Cell Design and Characterization}
\label{sec:cell}
\begin{figure}
\centering
\includegraphics[width=0.95\columnwidth]{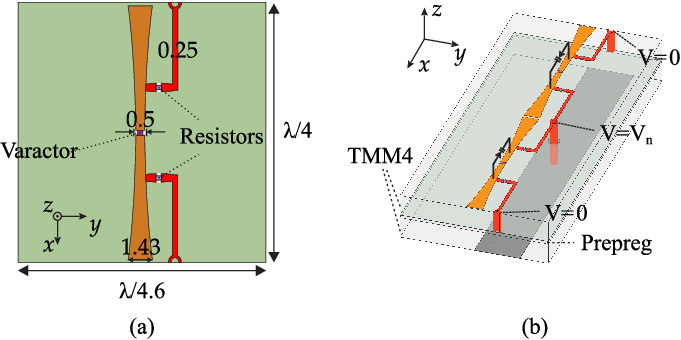}
\caption{\label{fig:Unit_cell} Tunable unit cell: (a) Top view. (b) Perspective view. A varactor is incorporated at each cell and it is reverse biased through a voltage line at the bottom layer and vias that connect to the voltage line and the ground plane.}
\end{figure}

The unit cell of the RIS is depicted in Fig.~\ref{fig:Unit_cell} on top and perspective views. The top layer takes the form of a tapered copper wire that is loaded with a varactor at its center and with a static capacitance at the edges (formed between the adjacent cells in the $x$-direction). The center frequency of operation is $5\mathrm{GHz}$, and the unit cell dimensions (in terms of center wavelength $\lambda$) are $\lambda/4 \times \lambda/4.6$. The subwavelength spacing in the $y$-direction is selected on purpose in order to be able to excite surface waves with high spectral $k_y$ components, that are necessary for beamshaping, as will become evident in the next section. A sparser design degrades the metrics of the achievable radiation patterns for some beamshaping examples. Similarly, a sparser periodicity in the $x$-direction compromises the accuracy of the homogenization of the loaded wires into impedance sheets, and decreases the tunability of the cell's surface impedance for a given range of the varactor's capacitance. The above aspects, together with the prototype's cost that increases with a higher density of cells, were considered to arrive at the unit cell's dimensions. The gallium arsenide MACOM MAVR-000120 varactor is employed in the design, providing a tunable capacitance from around $0.18 \mathrm{pF}$ (for a reverse bias voltage of $14\mathrm{V}$) to $1.2\mathrm{pF}$ (for zero bias voltage). The lumped resistance is selected to be $3 \Omega$ based on a previous fabricated RIS design. This approximate value will be validated experimentally in Sec.~\ref{subsec:Calibration}.  

The RIS stack-up comprises two Rogers TMM4 substrates ($\varepsilon_r=4.7$, $\mathrm{tan}(\delta)=0.002$) of thickness $h=1.9 \mathrm{mm}$ that are bonded together with the use of Panasonic R1650V prepreg of thickness $0.1 \mathrm{mm}$. Therefore, the total RIS thickness is approximately $h_\mathrm{tot} \approx 3.9\mathrm{mm}$ ($\approx 0.065\lambda$). The top impedance layer includes the varactor-loaded cells, the middle layer serves as a ground plane, whereas the bottom layer hosts the bias lines that carry the DC bias voltage. The biasing is achieved with vias that connect alternately at the ground plane and the bias line through an oversize hole in the ground plane. The polarity of the varactors in a single meta-wire also alternates so that all varactors are reverse-biased through the corresponding bias line in the bottom layer. Lastly, two $100 \mathrm{k\Omega}$ resistors are included to dissipate any RF currents induced on the bias lines.

\begin{figure}
\centering
\includegraphics[width=0.80\columnwidth]{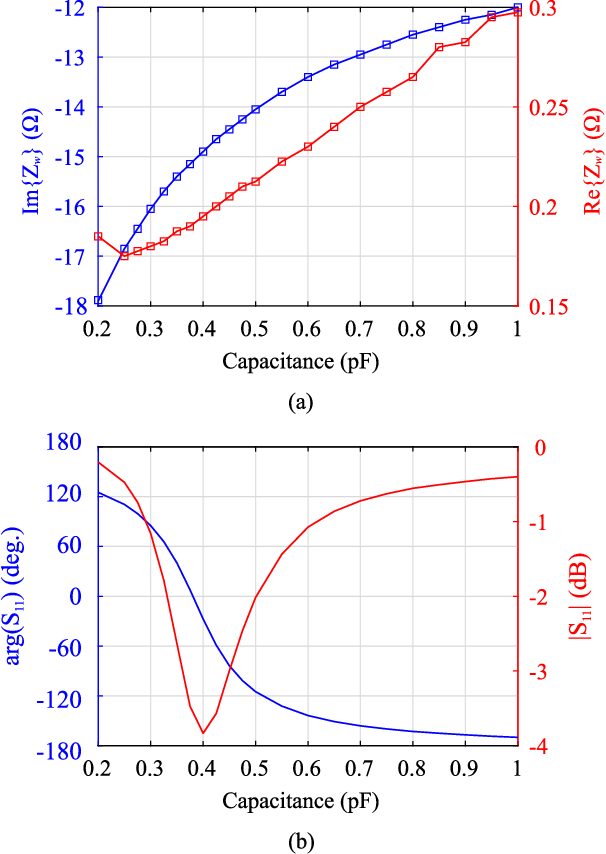}
\caption{\label{fig:Unit_cell_metrics}(a) Extracted surface impedance $Z_w$ as a function of the varactor's capacitance at $f=5 \mathrm{GHz}$. (b) Extracted reflection phase $S_{11}$ under periodic boundary conditions at $f=5 \mathrm{GHz}$.}
\end{figure}

The surface impedance of the unit cell as a function of the varactor's capacitance is first obtained, as described in \cite{Budhu:TAP2020,Xu:Access2022}. The finite RIS of length $L_\mathrm{tot}$ is simulated, but with the inclusion of only a single meta-wire (i.e., as if $N=1$). The meta-wire has the geometry of Fig.~\ref{fig:Unit_cell}, and it is placed at the center ($y=0$) of the impedance layer. The varactor's capacitance is varied and the scattered electric field from the single-cell RIS upon an incident illumination is recorded at a line segment of length $2\lambda$ that is located $\lambda/2$ above the RIS. Similarly, the scattered field from the homogenized RIS model, that features a single impedance strip of varying surface impedance $Z=R+jX$, is calculated at the same line segment through the MoM approach. For each capacitance, the impedance value $Z$ with the closest matching in terms of the scattered electric field values is assigned as the surface impedance. Compared to \cite{Budhu:TAP2020,Xu:Access2022}, the impedance is varied in a grid of real part and imaginary part values, allowing to additionally characterize the losses. The resulting graph of surface impedance as a function of capacitance is given in Fig.~\ref{fig:Unit_cell_metrics}(a). While the illumination (in this work, a normally-incident Gaussian beam) and the observation points for the scattered-field comparison are selected arbitrarily, the extracted values would not differ significantly for other choices. A higher accuracy may be achievable by accounting for the local field on each cell \cite{Budhu:TAP2023}. However, the matching between full-wave simulations and the homogenized-model predictions is already satisfactory (as will become evident in Sec.~\ref{sec:simulation}), and accounting for the local field would have required a different set of simulations for each cell and optimized solution.

Following a more typical approach, the unit cell is also characterized in terms of a reflection coefficient $S_{11}$ under periodic boundary conditions. The amplitude and phase of $S_{11}$ are given in Fig.~\ref{fig:Unit_cell_metrics}(b), allowing to establish a correspondence between surface impedances and reflection coefficients. While the surface impedances are utilized in the optimization framework, translating them to reflection coefficients is beneficial to predict the reflected radiation pattern through a simple reflect-array model and assess the benefits of the adopted integral-equation approach. 

\section{Simulation Results}
\label{sec:simulation}
The capabilities of the RIS for beamsteering and beamforming with high efficiency are examined in this section through full-wave simulations. The assumed RIS has a length of $L_\mathrm{tot}=6.5 \lambda \approx 390 \mathrm{mm}$ in the center frequency of $5 \mathrm{GHz}$. There are a total of $30$ meta-wires, each one with a surface impendace $Z_n$ to be optimized. The meta-wires take the form of the unit cell presented in Sec.~\ref{sec:cell} with their optimized surface impedances implemented through the corresponding capacitance values. Although the RIS is truncated to $L_x=4\lambda$ (16 cells) along the $x$-axis, the full-wave simulations assume an infinite structure by simulating a single row of unit cells and applying periodic boundary conditions along the $x$-axis. Nevertheless, the width $L_x=4\lambda$ is sufficient to maintain high-accuracy of the radiation pattern in the $yz$-plane, as will become evident from the measured results. 

For the MoM solution, during each optimization iteration, the induced currents are expanded to $N_g=600$, $N_w=150$ and $N_v=3000$ basis functions expanding throughout the ground plane, impedance wires and substrate, respectively. Moreover, the imaginary parts of the surface impedance, that constitute the optimization variables are constrained within the range $X_n \in [-17.85,-12.05] \Omega$ corresponding to a capacitance $C_n \in [0.2,1] \mathrm{pF}$, or a bias voltage $V \in [0.5,11.5]\mathrm{V}$. The stricter range compared to the nominal range of the varactor's model was selected so that any fabrication issues (e.g., shift in the substrate's permittivity) will not affect the realizability of the optimized surface impedance values.

\subsection{Beamsteering simulation results} \label{subsec:BeamSteering}
\begin{figure}
\centering
\includegraphics[width=0.96\columnwidth]{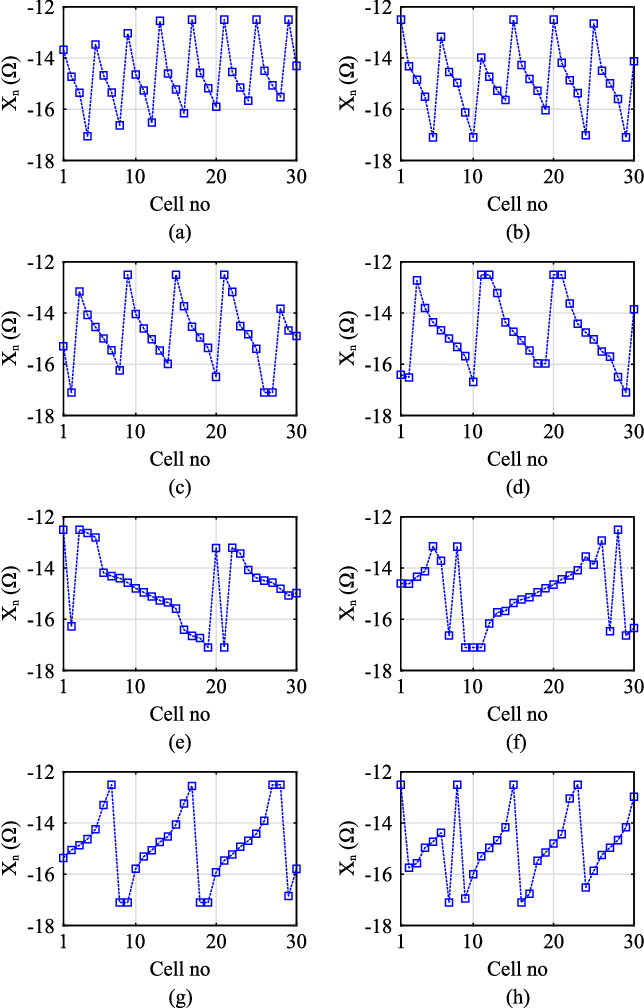}
\caption{\label{fig:Impedances_steering} Optimized surface impedance values $X_n=\mathrm{Im}\{Z_n\}$ for the $30$ meta-wires and for a desired reflected angle $\theta_\mathrm{out}$ of: (a) $-60^\circ$, (b) $-45^\circ$, (c) $-30^\circ$, (d) $-15^\circ$, (e) $0^\circ$, (f) $30^\circ$, (g) $45^\circ$, (h) $60^\circ$. }
\end{figure}

\begin{figure}
\centering
\includegraphics[width=0.80\columnwidth]{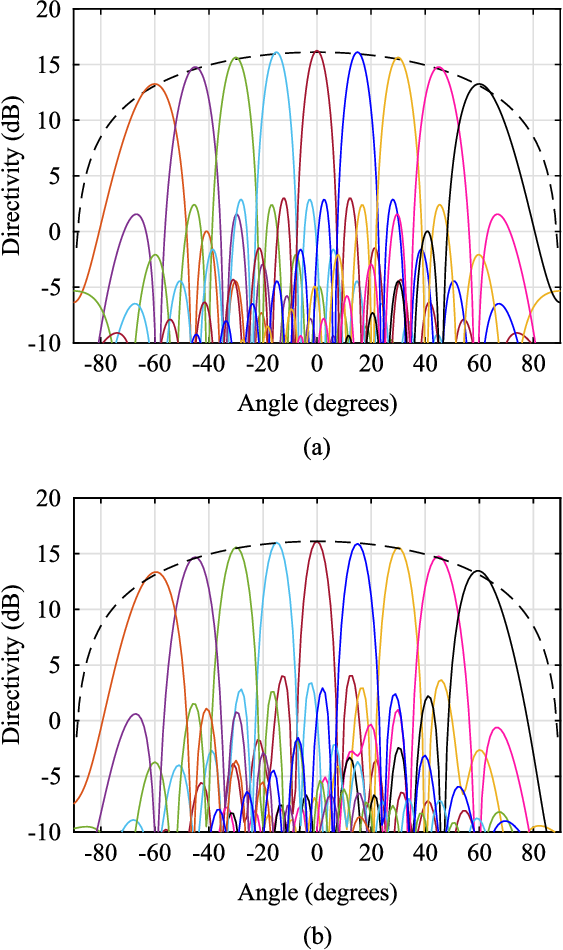}
\caption{\label{fig:Beamsteering_HFSS} Beamsteering demonstration of the RIS at $5 \mathrm{GHz}$: (a) Analytically-defined desired radiation patterns. (b) Predictions through full-wave simulations of the physical structure for each one of the optimized designs. The uniform-amplitude directivity approximation of Eq.~\eqref{eq:D_uni} is included with dashed line.}
\end{figure}

An incident plane wave at $f=5\mathrm{GHz}$ is impinging on the RIS from an angle of $\theta_\mathrm{in}=-15^\circ$. By optimizing the surface impedances through the presented framework, we explore the ability of the RIS to steer the beam with high directivity at the angular range of $\pm 60^\circ$ with a $15^\circ$ step. Although a $15^\circ$ step was selected for the reflected angle for presentation purposes, it is noted that beams with a maximum radiation towards any angle within this range can be realized. The specular case (reflection towards $15^\circ$) is not optimized, as the response can be directly obtained by uniformly loading the varactors at each meta-wire with the minimum capacitance value $C_n=0.2 \mathrm{pF}$. To calculate the desired far-field $E^{ff}_\mathrm{des} (\theta)$, an array of virtual sources is assumed along the RIS aperture with currents given as:
\begin{align}
I_{vs}[n]=\mathrm{exp}\{-jky_n \mathrm{sin}(\theta_\mathrm{out})\},
\end{align}
with $\theta_\mathrm{out}$ being the reflected angle for each case. The desired radiation patterns are plotted in Fig.~\ref{fig:Beamsteering_HFSS}(a). Optimization is then performed based on the cost function of Eq.~\eqref{eq:cost_function} to determine the required homogenized surface impedances. For each beamsteering case a power efficiency constraint $\eta_\mathrm{min}$ is selected, so that a satisfactory pattern matching can be achieved. The constraint values $\eta_\mathrm{min}$ for the various reflected angles ($\theta_\mathrm{out}$ is given in parenthesis) are: $0.51$~($-60^\circ$), $0.58$~($-45^\circ$), $0.6$~($-30^\circ$), $0.65$~($-15^\circ$), $0.62$~($0^\circ$), $0.62$~($30^\circ$), $0.6$~($45^\circ$) and $0.5$~($60^\circ$). The imaginary parts of the optimized surface impedances are given in Fig.~\ref{fig:Impedances_steering}(a)-(g) for all the reflected angles.

After the optimization stage, the surface impedances are implemented with unit cells of different capacitance based on Fig.~\ref{fig:Unit_cell_metrics}(a), and full-wave simulations of the physical structure are performed at Ansys HFSS. The obtained radiation patterns are plotted in Fig.~\ref{fig:Beamsteering_HFSS}(b), and they match closely with the desired patterns for all steering angles. For an aperture of size $L_\mathrm{tot}$ that exhibits a uniform-amplitude and linear phase to radiate towards an angle $\theta_\mathrm{out}$, the directivity can be approximated (in linear scale) as:
\begin{align} \label{eq:D_uni}
D_\mathrm{uni}=2\pi \left(\frac{L_\mathrm{tot}}{\lambda}\right) \mathrm{cos}(\theta_\mathrm{out}).
\end{align}
The expression in Eq.~\eqref{eq:D_uni} is also plotted in Fig.~\ref{fig:Beamsteering_HFSS}(a)-(b) for different angles $\theta_\mathrm{out}$ as an estimation of the maximum expected directivity. Moreover, an illumination efficiency can be defined as the ratio between the simulated directivity and the one for a uniform-amplitude aperture:
\begin{align} \label{eq:e_il}
\eta_\mathrm{il}=\frac{D(\theta=\theta_\mathrm{out})}{2\pi \left(\frac{L_\mathrm{tot}}{\lambda}\right) \mathrm{cos}(\theta_\mathrm{out})}.
\end{align}
The simulated values range from $\eta_\mathrm{il}=0.99$ to $1.08$ for the $9$ cases shown in Fig.~\ref{fig:Beamsteering_HFSS}(b), verifying that the realized beams are highly-directive in the $yz$-plane based on the RIS' aperture size. In addition, the sidelobe level is kept at $-11.2 \mathrm{dB}$ or lower for all cases (with an average of $-12.7 \mathrm{dB}$). Regarding the power efficiency, the values predicted by the integral-equation formulation for the converged solutions coincide in all cases with the imposed constraints. Furthermore, the power efficiency from the physical structure simulations ranges from $0.52$ to $0.63$, and it is within a $\pm 0.02$ range for all beamsteering cases, when compared to the imposed constraints. The above simulations verify the ability of the MoM framework to correctly predict the losses of the total structure during the optimization stage and reach solutions with satisfactory pattern matching, while avoiding excessive power losses. 

\subsection{Beamforming simulation results} \label{subsec:Beamforming}
\begin{figure}
\centering
\includegraphics[width=0.96\columnwidth]{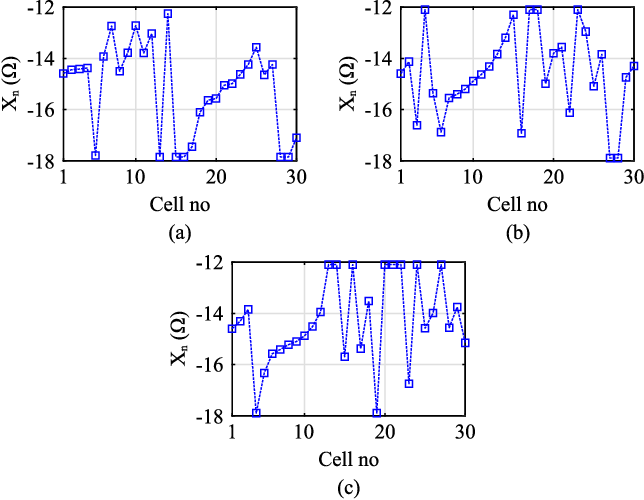}
\caption{\label{fig:Impedances_forming} Optimized surface impedance values $X_n=\mathrm{Im}\{Z_n\}$ for the $30$ meta-wires and for sector patterns centered at $\theta_\mathrm{out}=30^\circ$ and having a beamwidth of: (a) $\theta_b=30^\circ$, (b) $\theta_b=45^\circ$, and (c) $\theta_b=60^\circ$.}
\end{figure}

To demonstrate the beamforming capabilities of our framework and designed RIS, we assume again an incident plane wave from an angle of $\theta_\mathrm{in}=-15^\circ$. The desired patterns in this case are sector patterns that are centered at $\theta_\mathrm{out}=30^\circ$ and have a varying beamwidth $\theta_{b}=30^\circ, 45^\circ$ and $60^\circ$. The desired field pattern values are analytically defined for each case as:
\begin{align}
|E^{ff}_\mathrm{des} (\theta)|=\begin{cases}
1, & \theta \in [\theta_\mathrm{out}-\theta_b/2, \theta_\mathrm{out}+\theta_b/2], \\
0, & \mbox{elsewhere}.
\end{cases}
\end{align}
Such sharp sector patterns are impossible to realize with a finite RIS aperture, as there would always be a variation of the directivity within the beamwidth region, a smooth transition at the two limits and non-zero sidelobes. To facilitate optimized solutions with satisfactory metrics for these key characteristics the cost function in Eq.~\eqref{eq:cost_function} is supplemented with two extra terms, as follows:
\begin{align}\label{eq:cost_function2}
f'&=f + \left(\max_{\theta \in \theta_\mathrm{pass}} D_\mathrm{dB}(\theta)-\min_{\theta \in \theta_\mathrm{pass}} D_\mathrm{dB}(\theta) \right) \nonumber \\
&+ \mathrm{max}\left\{ \left(\max_{\theta \in \theta_\mathrm{SLL}}D_\mathrm{dB}(\theta) -\min_{\theta \in \theta_\mathrm{pass}} D_\mathrm{dB}(\theta)-C_\mathrm{SLL}\right),0\right\},
\end{align}
where $\theta_\mathrm{pass}$ and $\theta_\mathrm{SLL}$ are the angular ranges where the main sector beam and the sidelobes are expected, $D_\mathrm{dB}$ is the obtained directivity in dB scale and $C_\mathrm{SLL}$ is the maximum allowable sidelobe level (in dB). To allow for a realistic smooth transition, a ``buffer" range is allowed, and the afore-mentioned angular ranges are selected for the three cases as:
\begin{itemize}
\item $\theta_b=30^\circ$: \\
 $\theta_\mathrm{pass}=[17^\circ, 43^\circ]$,  $\theta_\mathrm{SLL}=[-90^\circ, 10^\circ] \cup [50^\circ, 90^\circ]$,
 \item $\theta_b=45^\circ$: \\
 $\theta_\mathrm{pass}=[10^\circ, 50^\circ]$,  $\theta_\mathrm{SLL}=[-90^\circ, 3^\circ]  \cup [57^\circ, 90^\circ]$,
 \item $\theta_b=60^\circ$: \\
 $\theta_\mathrm{pass}=[2^\circ,  58^\circ]$,  $\theta_\mathrm{SLL}=[-90^\circ, -6^\circ] \cup [66^\circ, 90^\circ]$.
\end{itemize}
The first additional term in Eq.~\eqref{eq:cost_function2} quantifies the variation of the directivity within the angular pass range. On the contrary, the second additional term acts like a penalty factor that is nonzero when the sidelobe level (obtained as the difference between the highest side lobe and the minimum directivity value within the pass range) exceeds the allowable SLL level that is set for all cases at $C_\mathrm{SLL}=-12\mathrm{dB}$. For these beamforming cases, the radiation pattern is calculated at a radius $r_d=2.5 \mathrm{m}$ during the optimization process, according to what described in Sec.~\ref{subsec:VSIE}. Although the difference compared to the radiation pattern in the far-field is generally small, the aim is to demonstrate the maximum possible accuracy compared to the measurements that are performed in a radius of $2.5 \mathrm{m}$ due to practical constraints. The optimization is performed with a power constraint $\eta_\mathrm{min}=0.5$ for all three designs and the optimized surface impedance values are given in Fig.~\ref{fig:Impedances_forming}.

\begin{figure}
\centering
\includegraphics[width=0.8\columnwidth]{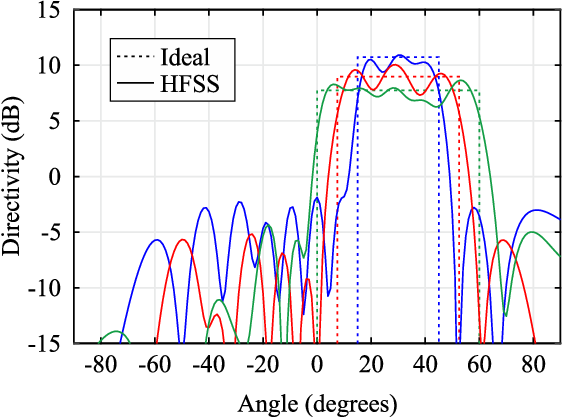}
\caption{\label{fig:Beamforming_HFSS} Beamforming demonstration of the RIS at $5 \mathrm{GHz}$ through $3$ sector patterns of beamwidth $\theta_b=30^\circ$ (blue color), $\theta_b=45^\circ$ (red color) and $\theta_b=60^\circ$ (green color). The ideal patterns (dashed lines) are shown alongside with the radiation patterns predicted from full-wave simulations (solid lines).}
\end{figure}

The corresponding desired radiation patterns for $\theta_b=30^\circ, 45^\circ, 60^\circ$ are plotted in Fig.~\ref{fig:Beamforming_HFSS} with dashed lines. On the other hand, the radiation patterns (at $r_d=2.5 \mathrm{m}$) obtained from full-wave simulations of the physical structure with the optimized capacitances are shown in Fig.~\ref{fig:Beamforming_HFSS} with solid lines. For all three cases, the sidelobes (within the range $\theta_\mathrm{SLL}$) are limited to $-12 \mathrm{dB}$ below the minimum value observed in the pass range $\theta_\mathrm{pass}$. On the other hand, the variation within the pass range is limited to $1.5 \mathrm{dB}, 2.7 \mathrm{dB}$ and $2.4 \mathrm{dB}$, respectively. Finally, the simulated power efficiency is $\eta_p=0.52, 0.46,0.51$ for the three designs, which is close to the constraint (and MoM prediction for the optimized solutions) of $\eta_p=0.50$.

\subsection{Surface-wave excitation and benefits of IE-MoM approach}
In this subsection, the accuracy of the proposed integral equation framework solved with a Method of Moments (IE-MoM) is compared with the traditional reflect-array (RA) approach that disregards any coupling effects. For the latter, a complex reflection coefficient $S^n_{11}$ is assigned to each one of the $30$ cells. By multiplying the incident field with the reflection coefficient, the reflected field can be found. Then, by sampling these values at the center $y=y_{cn}$, ($n=1,...,N$) of each unit cell an array factor is defined:
\begin{align} \label{eq:reflectarray}
AF_\mathrm{RIS}(\theta)=\sum_{n=1}^N \left(E^\mathrm{inc} (y=y_{cn}) \times S^n_{11} \right) \mathrm{exp}\{-jk_o r_n\},
\end{align}
with $r_n$ being the distance between the meta-wire $n$ and the observation point, and calculated as:
\begin{align} \label{eq:r_n}
r_n = \begin{cases}
-y_{cn} \mathrm{sin}(\theta), & r_d \to \infty, \\
\sqrt{(y_{cn}-r_d \mathrm{sin}(\theta))^2 +(h-r_d \mathrm{cos}(\theta))^2}, &\mathrm{finite } \  r_d.
\end{cases}
\end{align}
The directivity pattern calculated through the RA approach of Eqs.~\eqref{eq:reflectarray}-\eqref{eq:r_n} is compared against the prediction through the integral equation framework and with the full-wave simulations. In particular, the optimized solutions for beamsteering at $45^\circ$ (referring to the far-field) and for a sector pattern of $45^\circ$ beamwidth (refering to a radius of $r_d=2.5 \mathrm{m}$) are examined. The optimized surface impedances are directly used for the integral-equations framework, while they are also translated to the respective reflection coefficients through Fig.~\ref{fig:Unit_cell_metrics}(a)-(b), so that Eq.~\eqref{eq:reflectarray} can be used for the RA model.

\begin{figure}
\centering
\includegraphics[width=0.75\columnwidth]{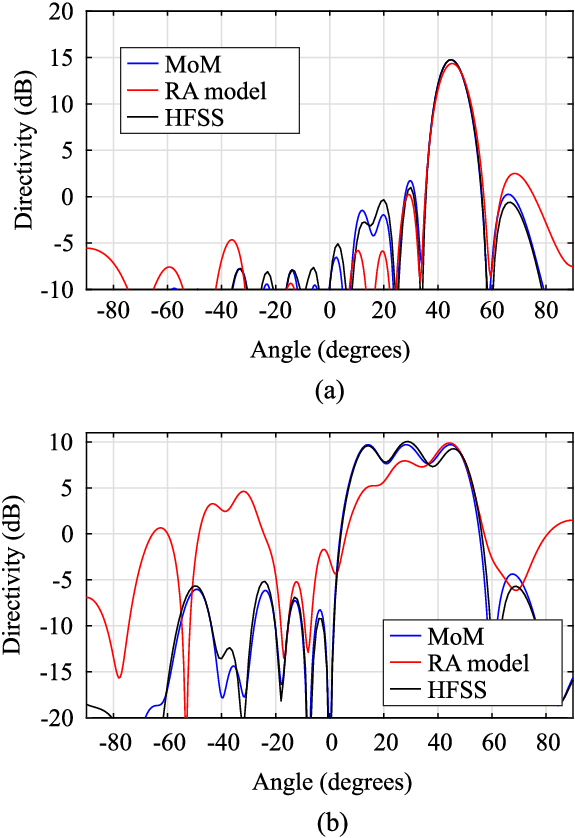}
\caption{\label{fig:Model_Accuracy} Predicted far-field radiation for: (a) Beamsteering at $45^\circ$, and (b) Beamforming to a sector pattern with $45^\circ$ beamwidth. The reflectarray (RA) model fails to provide high accuracy for the beamforming case, whereas the MoM solution matches closely with the full-wave simulations for all examined cases.}
\end{figure}

The radiation patterns for the two cases are plotted in Fig.~\ref{fig:Model_Accuracy}. It is evident that the RA approach predicts the pattern quite accurately for the beamsteering case, but significant discrepancies exist when beamforming and amplitude tapering are required, as is the case of realizing a sector pattern. On the other hand, the analysis of the structure through integral equations leads to results matching the full-wave simulations in both cases. Although two cases are presented for brevity, the RA model fails to predict accurately the pattern for all sector pattern solutions of Sec.~\ref{subsec:Beamforming}, whereas the discrepancies for the beamsteering cases of Sec.~\ref{subsec:BeamSteering} are insignificant.

\begin{figure}
\centering
\includegraphics[width=0.75\columnwidth]{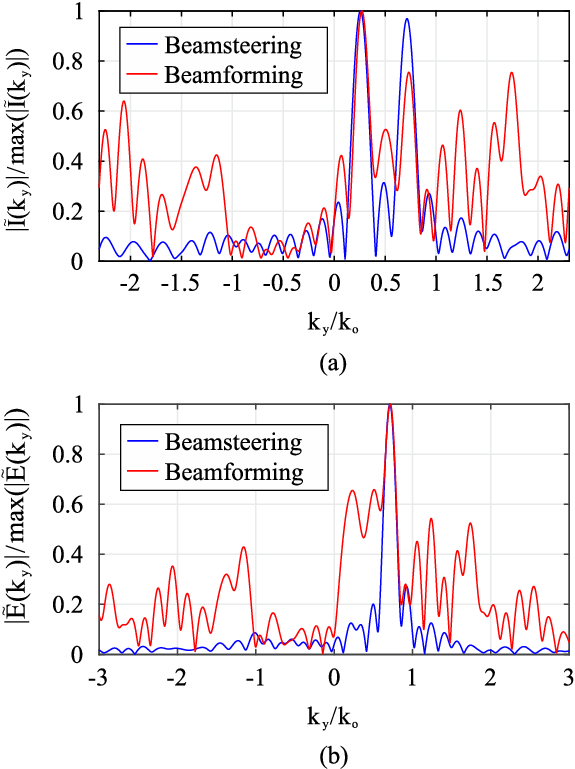}
\caption{\label{fig:Current_Field_Spectrum} Spectrum of: (a) the induced currents on the impedance layer, and (b) the scattered electric field at a plane $\lambda/20$ above the RIS. The cases of beamsteering at $45^\circ$ and beamforming into a sector pattern with a $45^\circ$ beamwidth are examined. The spectrum for the beamforming case exhibits higher-intensity surface waves that facilitate the amplitude tapering along the RIS aperture.}
\end{figure}

To further analyze the distinction between simple beamsteering and beamforming of the radiation pattern, the spectrum of the induced currents along the impedance strips is examined. For each of the strips, the predicted current density, as predicted by MoM, is integrated to calculate the total current $I_w[n]$. A continuous Fourier Transform is then calculated as:
\begin{align}
\tilde{I}(k_y)=\sum_{n=1}^N I_w[n] \mathrm{exp}\{+j k_y y_c[n]\}.
\end{align}
The current spectrum is depicted in Fig.~\ref{fig:Current_Field_Spectrum}(a) for the cases of beamsteering towards $+45^\circ$ and of beamforming into a sector pattern with $45^\circ$ beamwidth. As observed, the beamforming case presents much higher evanescent spectrum ($|k_y|>k_o$) for the induced currents. These induced fast-varying current components create corresponding surface waves in terms of the scattered electric fields, that are fostering amplitude tapering along the RIS aperture, as analyzed in previous works exploiting surface waves for beamforming \cite{Epstein:PRL2016,Salucci:TAP2018}. In particular, the spectrum of the scattered field for the two designed cases is plotted at a plane $\lambda/20$ above the RIS in Fig.~\ref{fig:Current_Field_Spectrum}(b). These surface waves are not captured properly by the simple RA model that treats the unit cells as isolated scatterers leading to discrepancies in the predicted scattered field. Additionally, the emergence of surface waves is typically accompanied with properties (e.g., surface impedance) that vary more quickly between consecutive cells. Therefore, the local periodicity approximation that assumes an infinite array in order to extract the reflected phase is less accurate for cases that high evanescent spectrum is excited. All the above highlight the advantages of the proposed integral-equation framework, that can  predict and optimize the radiation pattern more accurately at the expense of a slightly higher computational complexity.

\section{Measurements}
\label{sec:measured}

\subsection{RIS Prototype and Measurement Setup} 
\begin{figure}
\centering
\includegraphics[width=0.99\columnwidth]{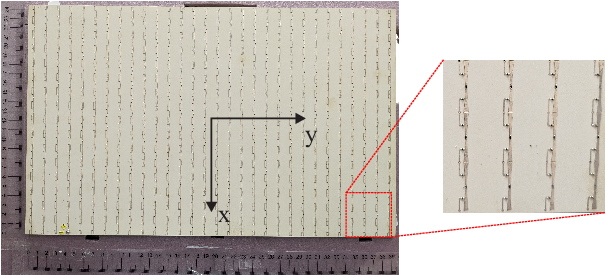}
\caption{\label{fig:RIS_prototype} Photograph of the top layer of the fabricated RIS prototype. Inset: Magnified photograph of the unit cells at the bottom-right corner.}
\end{figure}

An RIS prototype is fabricated based on the unit cell described in Sec.~\ref{sec:cell}. The RIS consists of $30$ meta-wires with $16$ unit cells each; thus, there are a total of $480$ unit cells. The total size of the RIS is around $6.5 \lambda \times 4\lambda$ (or $390 \mathrm{mm} \times 240\mathrm{mm}$). Two Measurement Computing Corporation (MCC) USB-3114 controllers are utilized to provide the $30$ analog output voltages to the voltage lines through connectors soldered in the back layer. A photograph of the RIS prototype is shown in Fig.~\ref{fig:RIS_prototype}. 

\begin{figure}
\centering
\includegraphics[width=0.95\columnwidth]{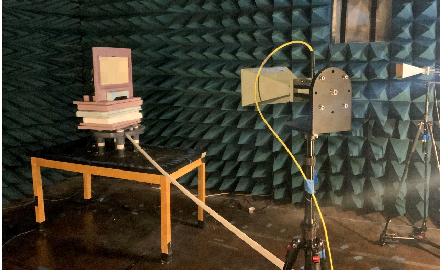}
\caption{\label{fig:Measurement_Setup} Bi-static measurement setup consisting of the RIS on the stage, the illuminating horn antenna at a fixed angle and the rotating receiving horn antenna.}
\end{figure}

The RIS is measured in a bi-static measurement setup located in an anechoic chamber, as shown in Fig.~\ref{fig:Measurement_Setup}. In particular, the RIS is placed on a stage and it is illuminated from the angle of $-15^\circ$ by a standard-gain horn antenna, that provides a gain of $16.8\mathrm{dB}$ at $5\mathrm{GHz}$ \cite{Ainfo:Horn15dB}. The distance of the transmitting antenna to the center of the RIS is $R_1=2.52 \mathrm{m}$ that guarantees that the incident field at the RIS closely resembles the plane-wave assumed in the optimization and simulations. On the other hand, an identical receiving horn antenna captures the reflected power from the RIS at different angles $\theta$ by rotating at a constant radius $R_2=2.5 \mathrm{m}$ from the RIS center. Measurements are taken through a VNA connected to both antennas and recording the $S_\mathrm{21} (\theta)$ coefficient. For the beamsteering case, measurements are taken every $1^\circ$ close to the expected peak (i.e., within a $\pm 20^\circ$ range) and every $2.5^\circ$ elsewhere. On the other hand, for the sector patterns demonstrating the beamforming capabilities, measurements are taken every $2.5^\circ$ for the whole pattern. The $S_{21}(\theta)$ values for each case are processed according to the procedure described in the Appendix~\ref{app:A} in order to extract the directivity pattern and the total, illumination and power efficiency of the RIS. 

A measurement in the absence of the RIS was performed to assess the noise floor of the setup. In this case, the measured power originates from reflections by non-absorptive surfaces in the chamber (e.g., table, floor) and from direct transmission between the antennas. For all angles, the noise floor was $29\mathrm{dB}$ to $32\mathrm{dB}$ below the measured peak of the beamsteering patterns, and $24\mathrm{dB}$ to $26\mathrm{dB}$ below the peak of the sector patterns. Therefore, the key characteristics of the radiation patterns (i.e., maximum directivity, beamwidth, major side lobes) are measured credibly, whereas some nulls or minor lobes may be affected by scattering unrelated to the RIS.

\subsection{Experimental Characterization of Varactors} \label{subsec:Calibration}
\begin{figure}
\centering
\includegraphics[width=0.8\columnwidth]{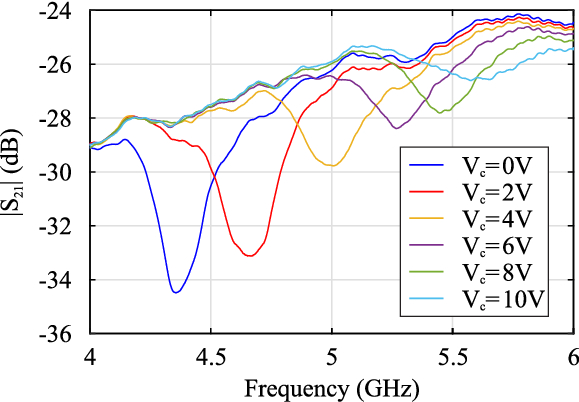}
\caption{\label{fig:Uniform_Specular} Specular reflection as a function of the common bias voltage for the uniformly-loaded RIS. The incident wave illuminates the RIS from $\theta=-15^\circ$.}
\end{figure}

\begin{figure}
\centering
\includegraphics[width=0.7\columnwidth]{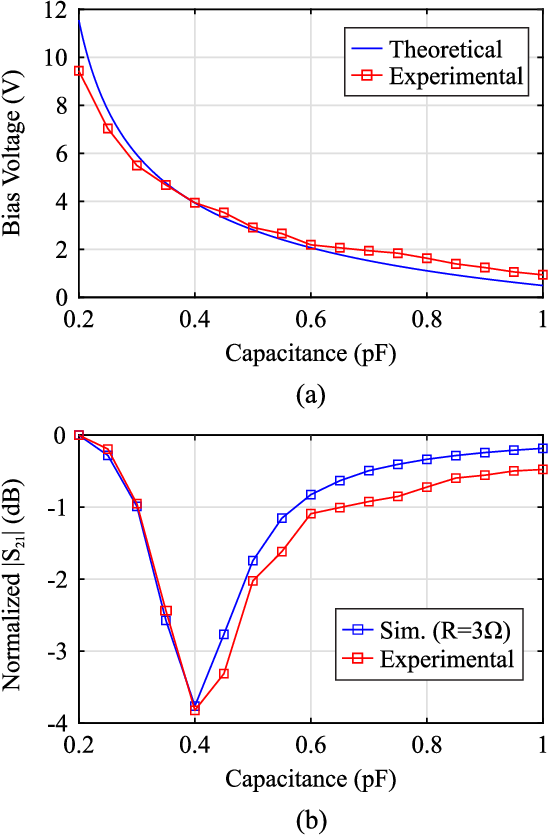}
\caption{\label{fig:Experimental_Char} Experimental characterization of the varactors through RIS measurements with uniform loading: (a) A capacitance-voltage (C-V) curve extracted from simulations is compared with the one extracted from the theoretical spice model. (b) The normalized losses at $5 \mathrm{GHz}$ as calculated from the measurements and from the unit cell's simulation with an $R=3 \Omega$ lumped resistance for each varactor.}
\end{figure}

It is paramount to characterize experimentally the varactors at $f=5 \mathrm{GHz}$, which is the intended frequency of operation for the RIS. This is both because the spice model from the manufacturer is not referring explicitly to the operating frequency and also because there is a variation of the varactors' capacitance for a given voltage due to manufacturing uncertainties. To extract the capacitance for a given voltage, all the bias voltage lines of the RIS are set to a common voltage level $V_c$. The specular reflection from an incident wave from $-15^\circ$ is measured for different levels of the common DC bias $V_c$ in the range of $0\mathrm{V}$ to $10\mathrm{V}$ with a step of $0.2 \mathrm{V}$. The measured RIS response as a function of frequency is shown for selected voltage levels in Fig.~\ref{fig:Uniform_Specular}, where it is evident that the resonance frequency shifts as a result of the change in the bias voltage and the corresponding capacitance of the varactors. By simulating a single unit cell in Ansys HFSS for different capacitance values in the range $[0.2,1] \mathrm{pF}$, the resonance frequency for different capacitances is determined. Then, through a comparison of the resonance frequency between the simulated data for different capacitances and the measured data for different voltages, a correspondance is established between the DC bias voltage and the effective capacitance. The relevant result is given in Fig.~\ref{fig:Experimental_Char} together with the theoretical curve extracted from the spice model and simulated in Keysight Advanced Design System (ADS). In general, the matching is satisfactory and the capacitance range of $[0.2,1]\mathrm{pF}$ is achieved within the limitations of the USB-controllers that provide an output voltage up to $10\mathrm{V}$. For all experiments, the experimentally-extracted curve in Fig.~\ref{fig:Experimental_Char}(a) is preferred to determine the required voltages from the optimized surface impedances/capacitances.

In addition to the capacitance value for different bias voltages, the assumed model of a series lumped resistance of $R=3\Omega$ for each varactor is verified through measurements. Specifically, the drop in the measured $S_{21}$ values at $5\mathrm{GHz}$ for different bias voltages $V_c$ (or, equivalently, for different capacitance values) is plotted in Fig.~\ref{fig:Experimental_Char}(b). In comparison, the simulated losses are plotted for an RIS unit cell under periodic boundary conditions, with an illumination from $-15^\circ$ and a lumped resistance of $R=3\Omega$ in series with the varactor's capacitance. Both curves are normalized to their peak value that is non-resonant and exhibits negligible losses. As observed, the series lumped resistance of $R=3\Omega$ is a reasonable estimation at $5\mathrm{GHz}$, as the curves match closely, especially for the capacitance values close to the resonance that exhibit the highest loss. It is highlighted, that both experimentally-extracted curves in Fig.~\ref{fig:Experimental_Char}(a)-(b) are approximate, as all other parameters of the RIS (e.g., permittivity or loss tangent of the substrate) are assumed to take their nominal values. Additionally, they only provide an average estimation of the varactors' parameters, as a small detuning may exist among the different varactors on the assembled RIS.

\subsection{Beamsteering Measurements}
\begin{figure}
\centering
\includegraphics[width=0.8\columnwidth]{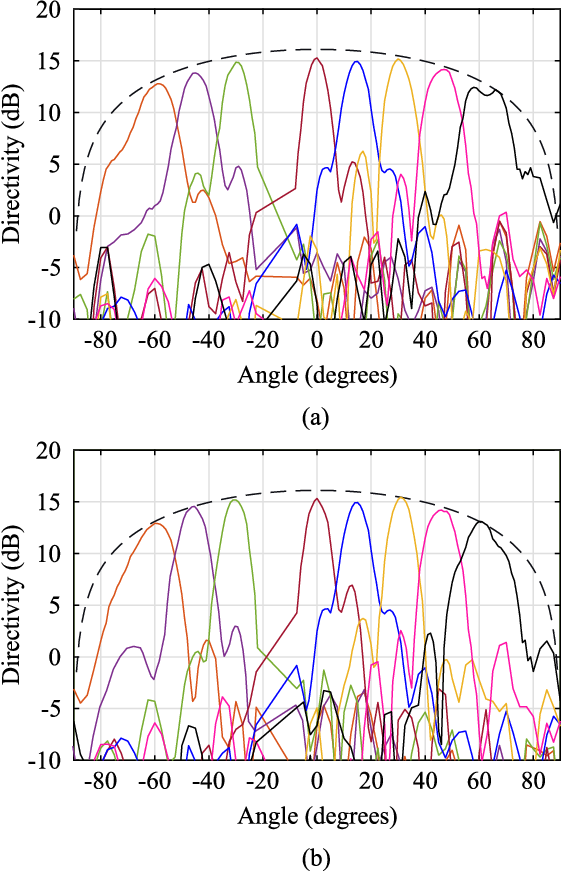}
\caption{\label{fig:Measured_BeamSteering} Measured directivity patterns for the beamsteering patterns at $5 \mathrm{GHz}$: (a) Patterns for the untuned sets of DC bias voltages that are optimized through the integral-equation framework. (b) Patterns for tuned sets of DC bias voltages. In both cases, the directivity of a uniform-amplitude aperture is included (dashed line).}
\end{figure}

The beamsteering patterns are first measured to assess the RIS ability to steer the incident beam with high directivity (limited only by the aperture size) and low power losses. As mentioned before, the frequency is set at $5\mathrm{GHz}$ and the illumination angle is $-15^\circ$. The optimized surface impedances of Fig.~\ref{fig:Impedances_steering} are converted to capacitances and applied DC bias voltages using the experimentally-extracted C-V curve. The set of $30$ voltage values are applied to the RIS meta-wires and the reflection is captured at different angles. The directivity patterns for these untuned cases are plotted in Fig.~\ref{fig:Measured_BeamSteering}(a). It is noted that the retro-reflection case ($\theta_\mathrm{out}=-15^\circ$) is not measured, as it is not possible to measure at a $\pm 7.5^\circ$ range around the transmitting antenna due to the two antennas colliding with each other. Moreover, the specular case is not optimized, but a measurement with all the voltages set to $0\mathrm{V}$ (producing a beam towards the specular) is included as reference. As observed, the patterns show an efficient steering of the reflected beam up to $\pm 60^\circ$ with the measured directivities for each case matching closely with the expected values in Fig.~\ref{fig:Beamsteering_HFSS}. The average illumination efficiency, as defined in Eq.~\eqref{eq:e_il}, is $0.87$ among the demonstrated steering angles and the average power efficiency is $0.61$, limited primarily by the varactors' losses.

In order to obtain the highest performance metrics from the RIS, a tuning step is performed for each set of voltages. In particular, the receiving antenna is positioned at the angle of the expected peak radiation and the voltages are tuned through particle swarm optimization at a tight range of $\pm 0.25 \mathrm{V}$ around their initial values. At every function evaluation, input is taken from the VNA regarding the $S_{21}$ transmission coefficient. Maximizing the received power through optimization leads to a higher total efficiency of the RIS for each beamsteering case. At the same time, the tight range is selected with the aim to maintain the nature of the initial optimized solution and only correct small issues stemming from the alignment of the setup, the variation of the individual varactors' properties, the finiteness of the RIS along the $x$-direction and any fabrication errors. The tuned patterns are given in Fig.~\ref{fig:Measured_BeamSteering}(b) and the relative metrics are included in Table~\ref{tab:BeamSteering} that summarizes all efficiency metrics both before and after tuning. On average, the tuning increases the average illumination efficiency to $0.94\%$ and the average power efficiency to $0.63$. While tuning provides a simple way to enhance the efficiency of the reflected beams for beamsteering, it is also evident that applying directly the untuned voltages leads to highly-directive beams that match well with simulations. The effectiveness of the untuned voltages is particularly crucial in cases that tuning cannot be applied easily, as in beamshaping cases of Sec.~\ref{subsec:Beamforming_measurements}.

Finally, the $3$-dB fractional gain bandwidth of the RIS for each case is ranging from $6.1\%$ to $9.8\%$ with an average value of $7.5\%$. The main limitation stems from the resonant behaviour of the unit cells that results in a frequency-dispersive homogenized impedance. It is noted that the reported values refer to the ``instantaneous" bandwidth for fixed sets of bias voltages optimized for the center frequency of $5 \mathrm{GHz}$. Moreover, the gain degradation is measured at the fixed desired angle, so beam-squinting effects also reduce the bandwidth.

\begin{table}
\caption{\label{tab:BeamSteering}Measured efficiency metrics for the tuned and untuned RIS beamsteering.}
\centering
\begin{tabular}{|c||l|l|l||l|l|l|}
\hline
 & \multicolumn{3}{|c||}{Untuned} & \multicolumn{3}{|c|}{Tuned} \\
\hline
Case & $\eta_\mathrm{il}$ & $\eta_p$ & $\eta_\mathrm{tot}$ & $\eta_\mathrm{il}$ & $\eta_p$ & $\eta_\mathrm{tot}$ \\
\hline
$-60^\circ$ & 0.81 & 0.52 & 0.42 & 1 & 0.54 & 0.54 \\
\hline
$-45^\circ$ & 0.94 & 0.57 & 0.53 & 0.93 & 0.59 & 0.55 \\
\hline
$-30^\circ$ & 0.93 & 0.67 & 0.62 & 1 & 0.67 & 0.67 \\
\hline
$0^\circ$ & 0.83 & 0.69 & 0.57 & 0.83 & 0.72 & 0.60 \\
\hline
$30^\circ$ & 0.87 & 0.62 & 0.54 & 0.93 & 0.68 & 0.63 \\
\hline
$45^\circ$ & 0.84 & 0.64 & 0.53 & 0.97 & 0.64 & 0.62 \\
\hline
$60^\circ$ & 0.90 & 0.58 & 0.52 & 0.96 & 0.54 & 0.52 \\
\hline
\textbf{Average} & \textbf{0.87} & \textbf{0.61} & \textbf{0.53} & \textbf{0.95} & \textbf{0.63} & \textbf{0.59} \\
\hline
\end{tabular}
\end{table}

\subsection{Beamforming Measurements}\label{subsec:Beamforming_measurements}
The beamforming capabilities of the RIS have been examined through full-wave simulations in Sec.~\ref{subsec:Beamforming}. Moreover, it was shown that the beamforming is enabled by auxiliary surface waves that are excited by high spectral components of the induced currents. These surface waves require an accurate framework to analyze the RIS response, such as the proposed analysis through integral equations, as a simple reflect-array model that disregards coupling effects is not sufficient to correctly predict the radiation pattern. In this section, the optimized sets of surface impedances to realize sector patterns of various beamwidths, as given in Fig.~\ref{fig:Impedances_forming}, are translated to DC bias voltages and measurements are performed. It is recalled that the frequency is $5 \mathrm{GHz}$ and the RIS is illuminated from an incident angle of $-15^\circ$. Unlike the beamsteering cases, an automated tuning is not possible in the sector-pattern cases, since there is not a single angle that we aim to maximize the reflected power. Hence, the radiation patterns for the solutions obtained directly through the proposed optimization framework are shown.

\begin{figure}
\centering
\includegraphics[width=0.75\columnwidth]{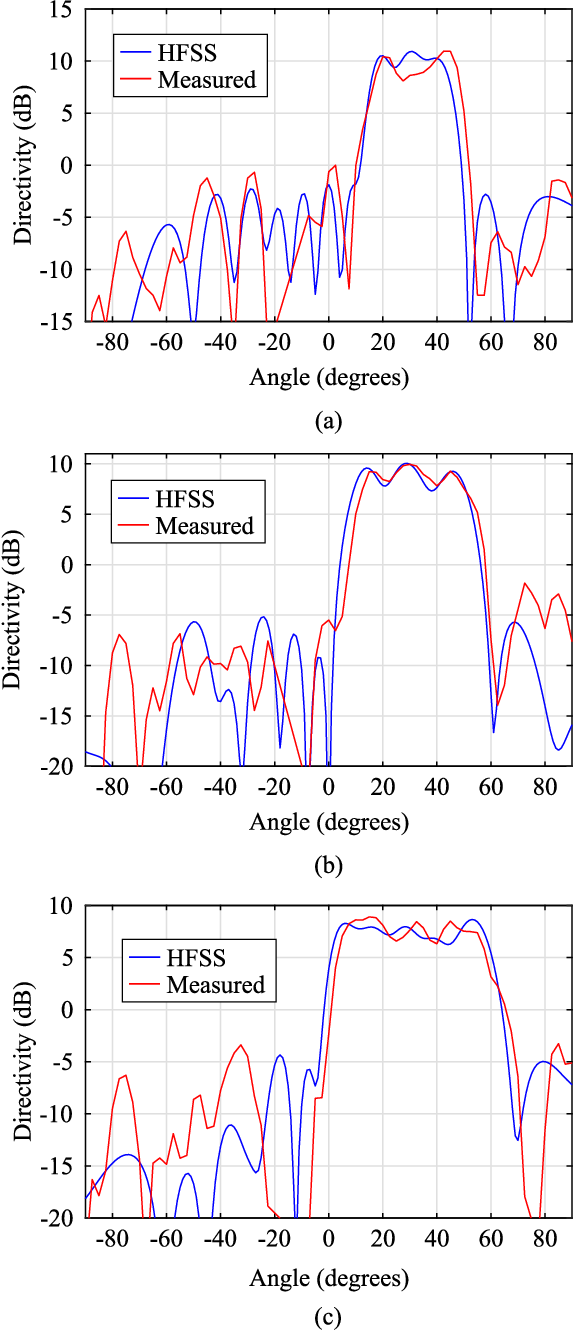}
\caption{\label{fig:Measured_BeamForming} Measured directivity patterns for the sector patterns at $5 \mathrm{GHz}$ with a beamwidth of: (a) $\theta_b=30^\circ$, (b) $\theta_b=45^\circ$, and (c) $\theta_b=60^\circ$. The patterns from full-wave simulations are also included for comparison.}
\end{figure}

The three measured directivity patterns for the three sector beams are plotted in Fig.~\ref{fig:Measured_BeamForming}(a)-(c) together with the simulated patterns. As observed the measured patterns match satisfactorily with the simulations, with the respective angular pass ranges identified as $\theta_\mathrm{pass}^\mathrm{m}=[-43^\circ,-16^\circ]$, $[-50^\circ,-12.5^\circ]$ and $\theta_\mathrm{pass}^\mathrm{m}=[-57.5^\circ,-4^\circ]$. These angular pass ranges are matching closely with the ones set as $\theta_\mathrm{pass}$ in Eq.~\eqref{eq:cost_function2} during optimization. Additionally, the directivity variation within these ranges is $2.9 \mathrm{dB}$, $2.5 \mathrm{dB}$ and $3 \mathrm{dB}$, respectively. Importantly, the sidelobe level (SLL) remains at least $-10.9 \mathrm{dB}$ compared to the peak value for all three sector patterns. The sidelobes in the beamforming case were severely affected in Fig.~\ref{fig:Model_Accuracy}(b), when the radiation pattern was predicted through the RA model. Similarly, a previous work that tried to optimize the RIS for realizing a sector pattern based on an RA model revealed issues with high SLL, probably due to the inaccuracies introduced by disregarding the coupling effects between adjacent unit cells \cite{Bagheri:Access2023}. Therefore, the results herein demonstrate that the key pattern characteristics, even for beamforming cases like the realized sector patterns, are accurately preserved through our high-accuracy optimization framework not only in full-wave simulations but also in the measured results. Lastly, the power efficiency is estimated to $0.51$, $0.52$ and $0.53$ for the three sector patterns, which are close to the expected power efficiency values from full-wave simulations.

\section{Conclusion}
\label{sec:conclusion}
In conclusion, the design of an RIS for beamsteering and beamforming at $5 \ \mathrm{GHz}$ was presented. The RIS consists of varactor-loaded unit cells that are represented with a tunable homogenized impedance. By optimizing the surface impedances through a rigorous integral-equation framework, the reflections of the RIS can be shaped into desired radiation patterns, while a constraint can be placed on the total power efficiency. Both the case of beamsteering to $\pm 60^\circ$ and the case of beamforming into sector patterns with varying beamwidth were investigated through simulations and experiments. For the beamforming case, it was shown that the subwavelength cells allow for the excitation of surface waves that are necessary to achieve both amplitude and phase tapering along the RIS aperture. Additionally, optimizing the RIS through the simplified integral-equation model shows high accuracy, when compared with full-wave simulations and a simplified reflectarray model that disregards any mutual coupling effects. 

The experiments verified beamsteering with directivities that approximate the theoretical limit of a uniform-aperture of the same size (i.e., average illumination efficiency of $95\%$) and with relatively high power efficiency averaging $63\%$ for steering up to $\pm 60^\circ$. On the other hand, the beamforming sector patterns displayed good matching with full-wave simulations. For all cases of varying beamwidth, the variation along the angular pass range was at or below $3\mathrm{dB}$, whereas the SLL was maintained below $-10.9 \mathrm{dB}$ and the estimated power efficiency was slightly over the constraint of $50\%$. Overall, the proposed work paves the way to realize RISs with enhanced beamforming capabilities that allow not only to steer the beam but fully control its characteristics with high fidelity.

\appendices
\section{Calculation of measured RIS efficiency} \label{app:A}
As mentioned in the main text, the receiving antenna records the $S_{21}(\theta)$ coefficient in different angles at a constant radius $R_2$. Given that $|S_{21}|^2$ is analogous to the power transmitted towards each angle $\theta$, the directivity pattern is directly calculated as:
\begin{align}
D_\mathrm{meas} (\theta)=2\pi \frac{|S_{21}|^2}{\int_{-\pi/2}^{\pi/2} |S_{21} (\theta)|^2 d\theta},
\end{align}
where the integration is performed numerically. Subsequently, the measured illumination efficiency $\eta_\mathrm{il}$ is calculated through Eq.~\eqref{eq:e_il} by using the measured directivity $D_\mathrm{meas}(\theta=\theta_\mathrm{out})$ as the nominator. The illumination efficiency quantifies how directive is the beam compared to a uniform-amplitude aperture of the same physical size that radiates towards $\theta_\mathrm{out}$. Of course, such a metric is relevant only in the beamsteering case that the aim is to maximize the received power at a single direction $\theta_\mathrm{out}$.

Furthermore, a total efficiency can be obtained. This is affected not only by the broadening of the beam in the $yz$-plane, but also by all other factors that limit the received power compared to an ideal lossless RIS reflecting towards the desired angle $\theta_\mathrm{out}$. The bi-static radar equation dictates that the transmission coefficient between the transmitting and receiving antenna ports is:
\begin{align} \label{eq:bistatic}
|S_{21}^\mathrm{ideal} (\theta_\mathrm{out},\theta_i)|^2=\frac{P_r}{P_i}=\sigma(\theta_\mathrm{out},\theta_i) \frac{G_{t}G_{r}}{4\pi} \left(\frac{\lambda}{4\pi R_1 R_2}\right)^2,
\end{align}
where $P_r,P_i$ are the received and incident power, respectively, $G_t, G_r$ are the gains of the transmitting and receiving antennas and $R_1, R_2$ are their distances from the center of the RIS. Finally, $\sigma(\theta_r,\theta_i)$ is the radar cross-section (RCS) of the RIS, which for an ideal case is \cite{Trichopoulos:OJCS2022}:
\begin{align} \label{eq:sigma_ideal}
\sigma(\theta_\mathrm{out},\theta_i)=4 \pi \frac{(L_x L_y)^2}{\lambda^2} \mathrm{cos}(\theta_i) \mathrm{cos}(\theta_\mathrm{out}),
\end{align}
where $L_x, L_y$ are the RIS' lengths along the $x$ and $y$ axis, respectively. By replacing the ideal RCS value of Eq.~\eqref{eq:sigma_ideal} into Eq.~\eqref{eq:bistatic}, an ideal $|S_{21}^\mathrm{ideal}|^2$ transmission coefficient can be calculated. Finally, the total efficiency is defined as the ratio between the measured and ideal values:
\begin{align}
\eta_\mathrm{tot}=\frac{|S_{21} (\theta_\mathrm{out})|^2}{|S_{21}^\mathrm{ideal} (\theta_\mathrm{out},\theta_i)|^2}.
\end{align}

Assuming that uniformity is maintained along the the $x$-axis, so that no excessive broadening exists in the elevation plane, the other factor that limits the total efficiency is the power losses of the RIS. Therefore, an estimation of the power efficiency for each beamsteering case is obtained as:
\begin{align}
\eta_p=\frac{\eta_\mathrm{tot}}{\eta_\mathrm{il}}.
\end{align}
Lastly, for the estimation of the power efficiency $\eta_p$ for the sector patterns, a slightly different approach is followed since there is not a single peak radiation angle. The expected power for a sector pattern realized through a lossless RIS is calculated over all angles in the $yz$-plane. In particular, the maximum transmission coefficient in Eq.~\eqref{eq:bistatic} is reduced by the difference between the maximum directivity of a directive beam and the directivity of an ideal sector-pattern beam, as given in Fig.~\ref{fig:Beamsteering_HFSS}(a) and Fig.~\ref{fig:Beamforming_HFSS}(a), respectively. By integrating both the expected $|S_{21}|^2$ values for the lossless RIS case and the measured values over all angles, the ratio that estimates the power efficiency is calculated.

\section*{Acknowledgement}
The work is supported by the Department of National Defence’s Innovation for Defence Excellence and Security (IDEaS) Program.



\end{document}